# The u-index: a simple metric to objectively measure academic impact of individual researchers

Roberto Dillon

James Cook University Singapore

School of Science and Technology

roberto.dillon@jcu.edu.au

ORCID: https://orcid.org/0000-0003-0166-0273

## Abstract

This short paper introduces the u-index ("unique-index"), a simple but robust metric to evaluate the impact and relevance of academic research output, as a possible alternative to common metrics such as the h-index or the i10-index. The proposed index is designed to address known issues with standard metrics such as inflated ratings resulting from self-citations or obtained thanks to extended co-authorship numbers, where every citation is attributed the same impact despite limited individual contributions. The new index makes also possible to differentiate scholars who would otherwise fall into the same h-index group, hence providing further insights into the actual impact of a specific individual researcher.

## 1. Introduction

Evaluating academic research performance in an objective way is a fundamental aspect of academic life that is central to hiring, promotions, grant evaluations etc. Metrics such the overall number of citations, the i10-index or the h-index, which was first introduced in 2005 [1] and quickly proved its reliability in identifying academic value across different fields and career stages [2], are common measures to evaluate academic performance and suitability for possible career advancement or awards.

Nonetheless, as also reported in different papers, including [3] and [4], these metrics, due to their extreme simplicity, also have significant inherent limitations as discussed in [5], [6] and [7]. For instance, the i10 and h-index eliminate lowly performing research outputs while also reducing the impact of truly relevant works that are referenced hundreds of times. Most importantly, by focusing on the quantity of average-performing scholarly work for a given scientist, they implicitly generate the need to have as many papers as possible with a minimum number of citations. This brought constant attempts to game the system by means of excessive self-citations and ever-growing co-authorship teams who then contribute to cite the group's work. With the importance of objective metrics growing within the academic world, there is a concrete risk for such tactics to become even more widespread to artificially increase rankings without necessarily proving the underlying research has any real academic merit or significance [8].

To overcome these possible ways to artificially inflate someone's index, several alternative indices have been proposed, including the e-index [9], to help differentiates scholars within the same h-index group, and the g-index [10], to emphasize the contribution of highly cited works.

## 2. The u-index

To accomplish a fine-grained classification of scholars who have the same h-index while also providing a more robust way to evaluate meaningful performance of individual authors in highly cited papers that contribute to the advancement of knowledge in their respective fields, a new index is proposed here named the "u-index" (for "unique-index") as defined in (1) for an individual paper:

$$\text{u-index} = \frac{I + S/2}{\sqrt{N}} \quad (1)$$

Where $I$ is the number of independent citations (i.e. citations by different authors not involved in the original paper), $S$ is the number of self-citation (i.e. citations by the same authors of the original paper) and $N$ is the number of authors included in the paper. Dividing by the square root of the number of authors, instead of simply dividing for the authors number, allows for a smoother tapering off for small groups while also still offering value in contributing to large cooperative projects.

For a specific author then the corresponding u-index would simply be the sum of the individual papers u-indices.

Using this formula, highly popular works would still be considered and even more so for works with a single author or a small group who are widely referenced by other scholars. Self-citations, while still considered as they are indeed an important and legit part of an ongoing research, have less weight. Finally, dividing by the square root of the number of authors rewards works by small groups closely working together while reducing the impact of large groups where individual contributions are, necessarily, less meaningful.

Sample u-index results for different papers are exemplified in Table 1:

| # | Total Citations | $I$ | $S$ | $N$ | u-Index |
|---|---|---|---|---|---|
| 1 | 100 | 95 | 5 | 1 | 97.5 |
| 2 | 100 | 95 | 5 | 3 | 56.3 |
| 3 | 100 | 80 | 20 | 10 | 28.5 |
| 4 | 10 | 8 | 2 | 2 | 6.4 |
| 5 | 10 | 2 | 8 | 2 | 4.2 |
| 6 | 500 | 10 | 490 | 20 | 57.0 |

Table 1: examples of how the u-index would evaluate different papers according to the corresponding number of authors and self-citations.

From Table 1 we can see how the u-index would evaluate the relevance and impact of papers with specific characteristics. A high-impact paper with 100 overall citations by a single author with only a moderate number of self-citations (5% of the overall number) would get a value close to the original number of citations. On the other hand, the same paper with three authors (row 2) would see the u-index value drop to 56.3. If the paper was written by a large group of ten researchers, with a more significant self-citation contribution worth 20% of the total number of citations (row 3), the u-index would drop to 28.5, offering a much smaller contribution to each of the authors involved. For an i10-index level paper by two authors (rows 4 and 5), we can also appreciate how the attempt to self-inflate the citation number would produce a u-index value that is 34% lower than a similar paper with a healthier ratio of independent versus self-citations. Last, in row 6, we can appreciate how the u-index would react to an extreme case of manipulation where a big group of twenty people would self-inflate their work to increase their overall citation number. In this case the u-index would be close to the one presented in row 2, despite the latter having only one fifth of the overall number of citations.

## 3. Conclusions

The u-index, despite its conceptual simplicity, can be a useful alternative to common metrics such as the total number of citations, the h-index and i10-index as it is more robust to possible attempts to artificially inflate those values while also including all research outputs and giving appropriate weight to top performing works. Different variations are also possible, depending on the desired type of analysis or comparison needed. For example an u10-index, obtained by adding together only the u-indices for the top 10 publications based on overall citation numbers, could also be adopted for assessing and comparing academic impact by limiting the analysis only to most representative and successful research outputs. Finally, the proposed metric can also help in providing more fine-grained individual insights compared to broader metrics, such as the h-index.